\documentclass[aps,prb,reprint,amsmath,amssymb,superscriptaddress,floatfix]{revtex4-2}
\usepackage{graphicx,xcolor,hyperref}
\usepackage{braket}
\usepackage[ignoreunlbld,norefs,nocites]{refcheck}
\begin{document}

\title{Emergence of Resonating Valence-Bond Correlations in Stretched Graphene}
\author{S.\ Azadi}
\email{sam.azadi@manchester.ac.uk}
\affiliation{Department of Physics and Astronomy, University of Manchester, Oxford Road, Manchester M13 9PL, United Kingdom}
\author{A.\ Principi}
\affiliation{Department of Physics and Astronomy, University of Manchester, Oxford Road, Manchester M13 9PL, United Kingdom}
\author{T. D. K\"{u}hne}
\affiliation{Center for Advanced Systems Understanding, Untermarkt 20, D-02826 G\"orlitz, Germany}
\affiliation{Helmholtz Zentrum Dresden-Rossendorf, Bautzner Landstra{\ss}e 400, D-01328 Dresden, Germany}
\affiliation{TU Dresden, Institute of Artificial Intelligence, Chair of Computational System Sciences, N\"othnitzer Stra{\ss}e 46 D-01187 Dresden, Germany}
\author{M.\ S.\ Bahramy}
\affiliation{Department of Physics and Astronomy, University of Manchester, Oxford Road, Manchester M13 9PL, United Kingdom}
\date{\today}

\begin{abstract}
Electronic correlations in graphene are generally considered weak due to the large bandwidth of its $\pi$ electrons. Here we show that tensile expansion of the honeycomb lattice provides a direct route to enhancing correlation effects. Using variational and diffusion quantum Monte Carlo, we compare a conventional Jastrow–Slater determinant wave function with a resonating-valence-bond (RVB) Jastrow–antisymmetrized geminal product ansatz for a series of stretched graphene lattices. We find that the energy gain of the RVB state relative to the single-determinant description increases with bond expansion up to a critical strain $\delta_{\mathrm{cr}}$, and decreases beyond it, revealing a nonmonotonic evolution of electronic correlations. The crossover is found to occur in the range $15\% < \delta_{\mathrm{cr}} < 20\%$, in agreement with mechanical stability limits. This behavior indicates a transition from a weakly correlated Dirac semimetal to a regime with enhanced non-dynamic correlation and short-range singlet pairing. Our results provide direct many-body evidence that lattice expansion drives graphene into a regime where RVB-like correlations become energetically favorable, offering a simple route to tuning correlation effects in Dirac materials.
\end{abstract}

\maketitle
Graphene provides a unique platform for exploring the interplay between lattice geometry and electronic correlation in low-dimensional systems \cite{Novoselov2004,Geim2007,Castro2009,Novoselov2005,Kotov2012,Sarma2011,Cao2018,Wagner2017}. In its equilibrium structure, the $\pi$-electrons form a wide-band Dirac semimetal, where the large hopping amplitude between neighboring $p_z$ orbitals leads to highly delocalized electronic states and relatively weak correlation effects. However, the strength of electronic correlations in such systems is not fixed, but can be tuned by modifying the balance between kinetic energy and Coulomb interactions. In particular, tensile expansion of the graphene lattice increases the C–C bond length, thereby reducing the orbital overlap and suppressing the hopping amplitude that controls the electronic bandwidth. Within this picture, lattice stretching effectively enhances the relative importance of electron–electron interactions and drives the system toward a more strongly correlated regime. Understanding how the electronic ground state evolves under such conditions is therefore of fundamental interest, as it provides insight into the emergence of correlation-driven phenomena in Dirac materials and related honeycomb lattices.

The effect of tensile strain on graphene \cite{Guinea2010} has been studied at the level of elasticity and band structure. First-principles calculations and experiments have established  that ideal monolayer graphene sustains large tensile deformation, with biaxial instability occurring at strains of order 15–20\% \cite{Verbiest2016,Rakshit2010,Lee2008}. Within tight-binding descriptions, strain primarily reduces the nearest-neighbor hopping amplitude, leading to modifications of the Dirac spectrum and a suppression of the electronic bandwidth \cite{Pereira2009,CastroNeto2009,Kotov2012}. This bandwidth reduction has been proposed as a route to enhancing interaction effects, for example by bringing van Hove singularities closer to the Fermi level and promoting correlation-driven instabilities \cite{Uchoa2007,McChesney2010,Pathak2010}. In parallel, the half-filled Hubbard model on the honeycomb lattice has provided a paradigmatic framework for understanding correlation effects in graphene, predicting a transition from a Dirac semimetal to an antiferromagnetic insulator with increasing effective interaction strength $U/t$ \cite{Paiva2005,Sorella2012,Assaad2013}. While these studies suggest that lattice expansion can drive graphene toward a strongly correlated regime, a direct parameter free \textit{ab-initio} many-body characterization of how electronic correlations evolve under controlled bond stretching remains largely unexplored.

Quantum Monte Carlo (QMC) methods \cite{Ceperley1977,Ceperley1986,Matthew2001,BeccaSorella} have become one of the most accurate and systematically improvable approaches for describing electronic correlation in real materials. In particular, variational and diffusion Monte Carlo calculations based on the conventional Slater–Jastrow determinant (JSD) wave function have demonstrated remarkable success in capturing correlation effects. The Jastrow factor efficiently incorporates short-range electron–electron and electron–nucleus correlations, substantially improving the description of many-body effects while preserving the fermionic antisymmetry through the Slater determinant. This framework has been successfully applied to a wide range of systems, including bulk solids \cite{Neil2004,Neil2005,Kolorenc2011,Foyevtsova2014,Azadi2025}, molecular structures \cite{Zen2018,Mazzola2014,sorella2007}, and low-dimensional materials \cite{Mostaani2015,Mostaani2017,spanu2009,Azadi2026}, providing highly accurate ground-state energies, cohesive energies, and structural properties. In particular, QMC calculations using JSD wave functions have proven effective for two-dimensional systems such as graphene and other layered materials, where reduced dimensionality enhances correlation effects.

However, the accuracy of a single SD description can deteriorate when the interatomic distance in a system increases. As the overlap between neighboring orbitals decreases, the electronic kinetic energy is reduced and the system approaches a regime where the static correlation effects become important. A typical manifestation of this behavior is observed in bond stretching, where the correct description of the electronic state requires a superposition of multiple configurations to account for the partial localization of electrons and the formation of correlated singlet pairs. Within QMC, these effects cannot be fully captured by the conventional JSD wave function. A more flexible alternative is provided by the antisymmetrized geminal product (AGP) ansatz, in which the wave function is constructed from correlated electron pairs rather than independent single-particle orbitals. When combined with a Jastrow factor, the resulting Jastrow–AGP (JAGP) form naturally incorporates multi-determinant character and provides an efficient description of static correlation and resonating valence-bond (RVB) \cite{Pauling,Anderson1987,Capriotti2007,AndersonPRL87,Baskaran2003,Kivelson2003,Schaffer2007,OzoneTurboRVB} like pairing correlations in extended systems.

This work deploys real-space variational quantum Monte Carlo (VMC) and diffusion quantum Monte Carlo (DMC) methods \cite{TurboRVB,Marchi2009,Marchi2011,ccECP1,ccECP2,lda,Azadi2010,Fahy90,Umrigar2007,Sorella98,Casula2006,sorella2005} combined with
density functional theory (DFT) \cite{QE,QE2,PBE} to elucidate the effect of isotropic biaxial strain on the many-body Coulomb interaction between electrons and single-particle spectra of graphene. We investigate the evolution of electronic correlations in graphene under controlled bond expansion.  In VMC, expectation values are computed by sampling a parametrized trial wave function in real space, JSD and JAGP in this work, and optimizing its parameters to minimize the variational energy. The accuracy of VMC is therefore determined by the flexibility of the chosen ans\"{a}tz and the quality of its nodal structure. DMC goes beyond this by projecting the trial wave function onto the ground state through an imaginary-time evolution of the Schr\"{o}dinger equation, effectively filtering out higher-energy components. In practice, DMC is implemented with the fixed-node approximation\cite{Mitas2006}, which constrains the nodal surface to that of the trial wave function, making the resulting energy an upper bound to the exact ground-state energy within this constraint. Details of DFT and QMC calculations are provided in Supplementary Materials \cite{suppl}.

By comparing the energies obtained with a JSD wave function and a JAGP ans\"{a}tz, we directly quantify the role of static many-body correlation embedded in JAGP as a function of isotropic biaxial stretch. We find that the energy gain of the JAGP wave function increases with bond stretching up to a critical strain $15\% < \delta_{\mathrm{cr}} < 20\%$, and decreases beyond it, revealing a nonmonotonic evolution of correlation effects. These results indicate that tensile expansion drives graphene from a weakly correlated Dirac semimetal into a regime with enhanced RVB-like correlations, with maximal pairing character at intermediate bond expansion.

\begin{figure}
    \centering
    \includegraphics[width=1.0\linewidth]{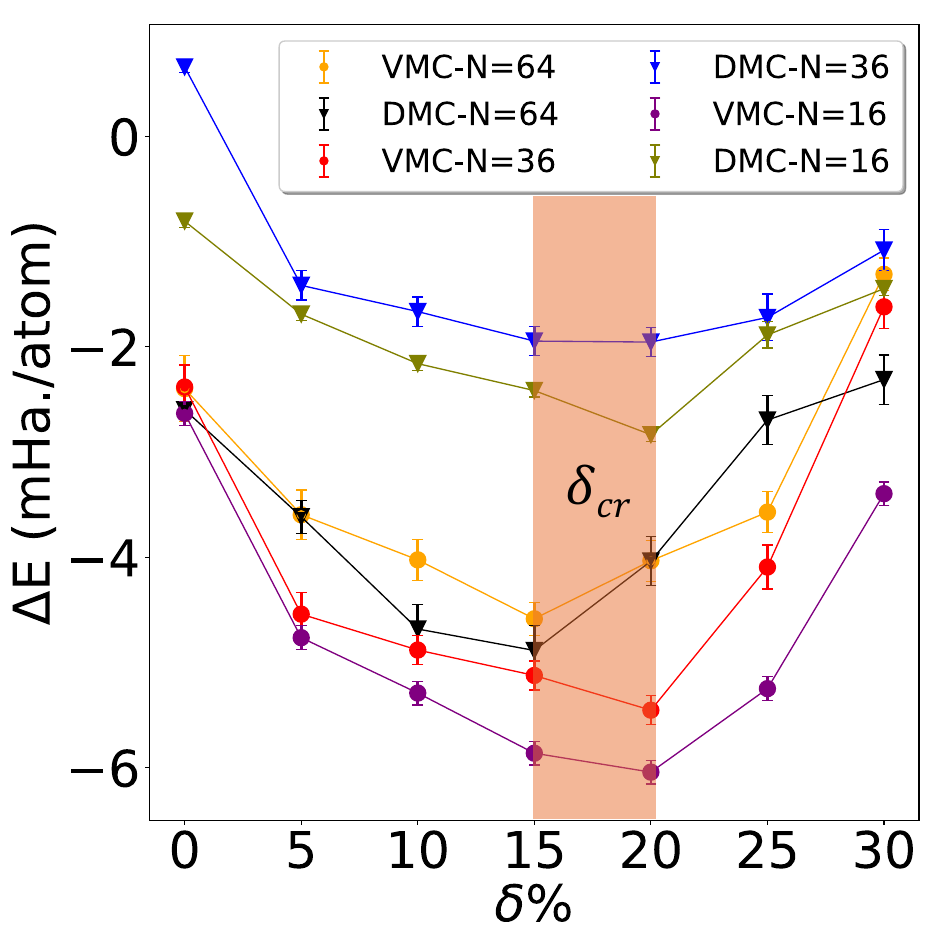}
    \caption{Energy difference $\Delta E = E_{\mathrm{JAGP}} - E_{\mathrm{JSD}}$ as a function of in-plane biaxial strain $\delta$, obtained from VMC and DMC calculations for system sizes $N$ = 16, 36, and 64 atoms. The shaded region indicates the critical strain $\delta_{\mathrm{cr}}$, where $\Delta E$ reaches its minimum; for $\delta > \delta_{\mathrm{cr}}$, $\Delta E$ increases with further stretching. The nonmonotonic behavior reflects the enhancement of static, RVB-like singlet correlations up to $\delta_{\mathrm{cr}}$, followed by a crossover to a different correlation regime at larger strain.}
    \label{fig:Evsdelta}
\end{figure}

The VMC and DMC energies obtained with both JSD and JAGP wave functions for system sizes of N = 16, 36, and 64 atoms show a systematic increase in total energy with increasing lattice expansion $\delta$ (Fig.~\ref{fig:Evsdelta}). As expected, the equilibrium configuration at $\delta = 0$ corresponds to the minimum energy for all system sizes and wave function ans\"{a}tze. The energy difference between JAGP and JSD wave functions provides a direct measure of the pairing component of the electronic correlation energy. The observed increase in the magnitude of $E_{\mathrm{JAGP}}-E_{\mathrm{JSD}}$ with bond expansion up to $15-20\%$ indicates  RVB–like singlet correlations. This behavior is consistent with a bandwidth-controlled increase of static correlation in the honeycomb lattice. However, the subsequent reduction of the JAGP energy gain at larger expansion signals a qualitative change in the nature of the correlated ground state. This nonmonotonic behavior suggests a crossover from an itinerant regime with enhanced singlet pairing correlations to a regime in which localization or alternative correlation mechanisms become dominant. Our results therefore demonstrate that tensile expansion does not simply strengthen electronic correlations, but drives a transition between distinct correlated regimes, with RVB-like correlations maximized at intermediate bond lengths.

The magnitude of the energy gain obtained with the JAGP wave function relative to JSD, in both VMC and DMC calculations, increases with lattice expansion up to a critical strain $\delta_{\mathrm{cr}} \approx 15\%–20\%$, after which it decreases. This trend indicates that tensile expansion enhances the importance of static, multi-configurational correlation, which is efficiently captured by the singlet pairing structure of the JAGP ansatz. The fact that $\delta_{\mathrm{cr}}$ lies close to the mechanical instability threshold of graphene suggests that these correlations are maximized in the last mechanically stable regime of the lattice. In this region, the reduction of orbital overlap and the resulting near-degeneracy of electronic configurations render a single-determinant description increasingly inadequate, while correlated singlet pairing provides a more favorable description of the ground state. These results therefore indicate that electronic correlation plays a significant role in stabilizing the stretched bonded state, lowering its energy relative to a single-determinant picture, and that the onset of mechanical instability occurs in a regime where many-body effects are already strongly enhanced.

The DFT electronic structure of graphene is characterized by a clear separation between $\pi$ and $\sigma$ bands \cite{suppl}. The bands forming the Dirac cones at the K points originate from $p_z$-derived $\pi$ and $\pi^*$ states, while the saddle point at M is also associated with the same $\pi$-band. In contrast, states near the $\Gamma$ point are dominated by in-plane $\sigma$ and $\sigma^*$ bands derived from $sp^2$ orbitals, with a gap separating valence and conduction branches. Along the $\Gamma$–K and $\Gamma$–M directions, the $\pi^*$ and $\sigma^*$ conduction bands can cross due to their distinct symmetry under mirror reflection with respect to the graphene plane, which suppresses hybridization between out-of-plane and in-plane states. 

The DFT band structures under isotropic in-plane tensile expansion show that the primary effect of strain is a strong renormalization of the conduction manifold, while the low-energy Dirac physics remains intact \cite{suppl}. The degeneracy among conduction bands is lifted by $\delta$, and the conduction $\pi^*$ and $\sigma^*$ bands shift toward the Fermi level as the C–C bond length increases. In contrast, the Dirac crossing at the K point is preserved, consistent with the maintained lattice symmetry under biaxial strain. A flattening of the $\pi^*$ band is observed prior to its crossing with the $\sigma^*$ band along the $\Gamma$–K and $\Gamma$–M directions, indicating a reduction of the effective kinetic energy scale. In addition, the bonding–antibonding splitting within the in-plane $\sigma$ bands decreases, leading to a reduction of the valence–conduction gap at the $\Gamma$ point. The DFT band structure results indicate a systematic suppression of orbital overlap and bandwidth under increasing $\delta$.

These single-particle trends provide a direct interpretation of the QMC results. The flattening of the $\pi^*$ band and the reduction of antibonding splittings imply a decrease in the hopping amplitude and, hence, an increase in the effective interaction-to-bandwidth ratio. This is consistent with the observed increase in the energy gain of the JAGP wave function relative to JSD up to intermediate strain, indicating the growing importance of non-dynamic correlation and short-range singlet pairing. The subsequent reduction of this energy gain at larger expansion correlates with the continued shift and restructuring of the conduction bands, suggesting a crossover from a regime dominated by delocalized RVB-like correlations to one where increased localization and changes in the orbital character of the low-energy excitations modify the nature of the correlated ground state.

The projected density of states (pDOS) provides a consistent spectral interpretation of the strain-dependent band structure \cite{suppl}. The $p_z$-resolved DOS shows that the van Hove singularities associated with the $\pi$ bands are preserved under tensile expansion, but become narrower and move closer in energy, reflecting a reduction of bandwidth and band flattening without opening a gap at the Fermi level. At the same time, the $s+p_x+p_y$–projected DOS reveals that the antibonding $\sigma^*$ states shift toward lower energies and gain weight above the Fermi level, reducing the bonding–antibonding splitting within the $\sigma$ band. This leads to a decrease of the $\sigma-\sigma^*$ gap at $\Gamma$ and an increasing energetic proximity between $\pi^*$ and $\sigma^*$ states, in agreement with the band structure. The DOS confirms that tensile strain primarily suppresses the kinetic energy scale and compresses the conduction spectrum while preserving the gapless Dirac character of graphene.

Although increasing the lattice expansion $\delta$ enhances static correlation in graphene, particularly at intermediate C–C bond lengths, this effect cannot be mapped directly onto the single-parameter Hubbard model on the honeycomb lattice. In the latter, increasing $U/t$ drives a well-defined transition from a Dirac semimetal to an antiferromagnetic insulating state, with no stable spin-liquid phase\cite{Sorella1992,Sorella2012}. In contrast, tensile expansion in graphene modifies not only the effective hopping amplitude but also the full band structure and orbital energetics. As a result, $\delta$ does not provide a one-to-one correspondence with U/t, and the evolution of correlation in the stretched lattice cannot be interpreted within a simple Hubbard framework. The present results therefore highlight that, while the honeycomb Hubbard model offers useful qualitative insight, a quantitative connection between lattice models and first-principles many-body calculations requires a more detailed mapping that accounts for multi-orbital effects and long-range interactions.

In this work, we have investigated the evolution of electronic correlations in graphene under isotropic tensile expansion using a combination of DFT and QMC methods. The DFT band structures and pDOS show that strain preserves the gapless Dirac spectrum while strongly renormalizing the conduction band, leading to band flattening, reduced antibonding splittings, and a compression of the energy scale associated with both $\pi^*$ and $\sigma^*$ states. These single-particle effects indicate a systematic reduction of the kinetic energy scale with increasing C–C bond length.

The QMC results provide direct many-body evidence of how these changes impact electronic correlation. The increasing energy gain of the JAGP wave function relative to the JSD up to intermediate strain demonstrates the growing importance of static, multi-configurational correlation and short-range singlet pairing. The subsequent reduction of this gain at larger expansion reveals a nonmonotonic evolution of correlation effects, indicating a crossover between distinct correlation regimes. These findings show that tensile expansion does not simply enhance correlation in a uniform manner, but instead drives the system toward a regime where RVB–like correlations are maximized before further evolution toward a different, more localized correlated state.

We acknowledge the support of the Leverhulme Trust under the grant agreement RPG-2023-253. S. Azadi and T.D. K\"{u}hne acknowledge the computing time provided to them on the high-performance computers Noctua2 at the NHR Center in Paderborn (PC2). The data that support the findings of this article are openly available \cite{github}, embargo periods may apply.

\bibliography{main}

{\Large \textbf{Supplemental Materials}}

\section{DFT simulations}
The DFT calculations were performed using the Quantum ESPRESSO package~\cite{QE}, employing the Perdew–Burke–Ernzerhof (PBE) exchange–correlation functional~\cite{PBE}. A plane-wave kinetic energy cutoff of 100 Ry and a charge density cutoff of 1200 Ry were used. Ultrasoft pseudopotentials~\cite{QE2} with four valence electrons were adopted for carbon. The band structures and projected densities of states were computed using a $24 \times 24 \times 1$ k-point mesh. To eliminate spurious interactions between periodic images, a vacuum spacing of $20 \AA$ was introduced along the out-of-plane $z$ direction.
\begin{figure*}
    \centering
    \includegraphics[width=1.0\linewidth]{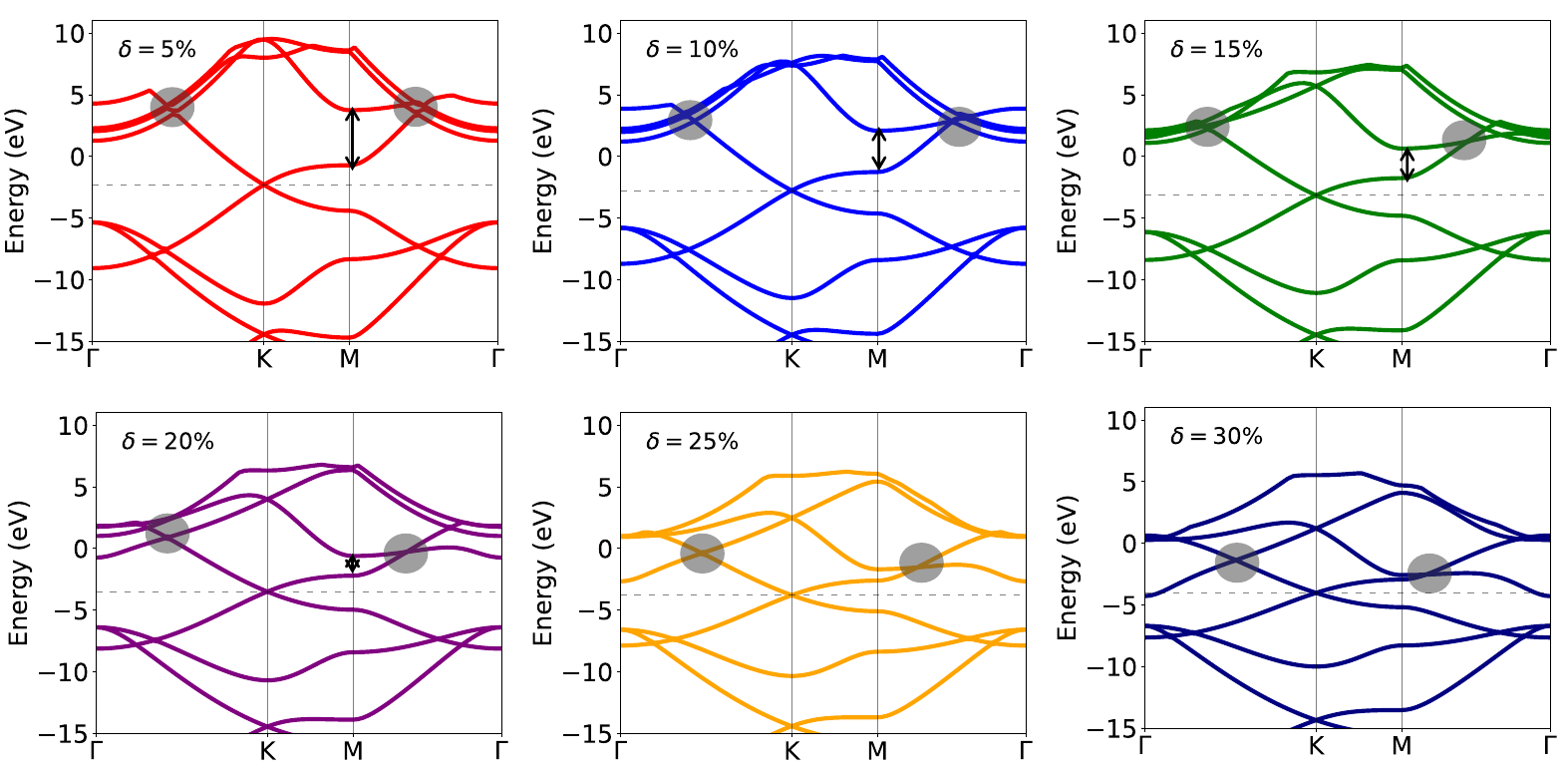}
    \caption{DFT band structure of graphene obtained at biaxial stretch $\delta=5\%, 10\%, 15\%, 15\%, 20\%,25\%,30\%$. The shaded regions in the conduction bands along the $M\text{–}\Gamma$ and $K\text{–}\Gamma$ directions highlight the crossings between the $\pi^*$ and $\sigma^*$ bands. The evolution of these crossings with increasing $\delta$ is shown. The arrow at the M point indicates the gap between the two lowest conduction bands, which decreases with increasing $\delta$ due to the reduced slope  and flattening of the $\pi^*$ band.}
    \label{fig:DFTBand}
\end{figure*}
\begin{figure*}
    \centering
    \includegraphics[width=1.0\linewidth]{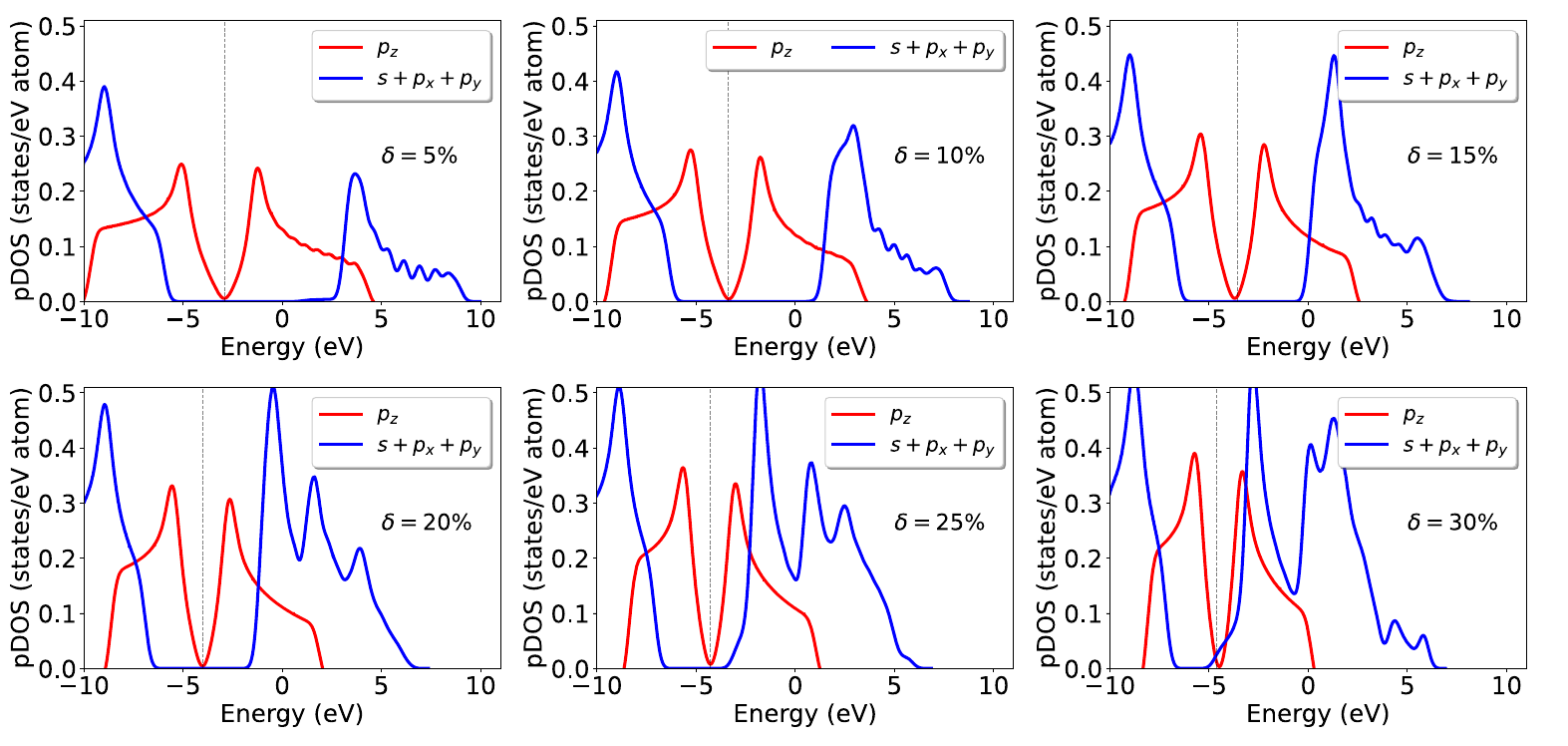}
    \caption{DFT project density of states (pDOS) of graphene obtained at biaxial stretch $\delta=5\%, 10\%, 15\%, 15\%, 20\%,25\%,30\%$.}
    \label{fig:DOS}
\end{figure*}

\section{QMC simulations}
The RVB wave function used in this work is defined as
\begin{equation}
\ket{\Phi_{\mathrm{RVB}}} = J(\mathbf{R}) \ket{\Psi_{\mathrm{AGP}}},
\end{equation}
where $J(\mathbf{R})$ is the Jastrow factor accounting for short-range electron–electron correlations, and $\ket{\Psi_{\mathrm{AGP}}}$ is the antisymmetrized geminal power (AGP) wave function given by
\begin{equation}
\Psi_{\mathrm{AGP}}(\mathbf{R}) = \mathcal{A} \prod_{i=1}^{N_{\downarrow}} \phi(\mathbf{r}_i^{\uparrow}, \mathbf{r}_i^{\downarrow}),
\end{equation}
where $\mathcal{A}$ is the antisymmetrization operator, $\mathbf{R} = \{ \mathbf{r}_1^{\uparrow}, \ldots, \mathbf{r}_{N_{\uparrow}}^{\uparrow}, \mathbf{r}_1^{\downarrow}, \ldots, \mathbf{r}_{N_{\downarrow}}^{\downarrow} \}$ is the set of electronic coordinates, and $\phi(\mathbf{r}_i^{\uparrow}, \mathbf{r}_i^{\downarrow}) = \phi(\mathbf{r}_i^{\downarrow}, \mathbf{r}_i^{\uparrow})$ is a symmetric pairing function describing singlet electron pairs.

The pairing function is expanded in a molecular orbital (MO) basis as
\begin{equation}
\phi(\mathbf{r}^{\uparrow}, \mathbf{r}^{\downarrow}) = \sum_{i=1}^{M} \alpha_i \phi_i^{\mathrm{MO}}(\mathbf{r}^{\uparrow}) \phi_i^{\mathrm{MO}}(\mathbf{r}^{\downarrow}),
\end{equation}
where $M \geq N_{\mathrm{el}}/2$, and each molecular orbital is expressed in a Gaussian basis $\{ \chi_j \}$ centered on the atomic positions,
\begin{equation}
\phi_i^{\mathrm{MO}}(\mathbf{r}) = \sum_j \beta_{ij} \chi_j(\mathbf{r}).
\end{equation}
The variational freedom increases for $M > N_{\mathrm{el}}/2$, while the minimal case $M = N_{\mathrm{el}}/2$ reduces to the Jastrow–Slater determinant (JSD) wave function~\cite{BeccaSorella}. In this work, we restrict to singlet states, so that the pairing function is real and symmetric. The number of MOs is chosen as the minimum required to reproduce a product of independent Hartree–Fock atomic wave functions~\cite{Marchi2009,Marchi2011}. The Gaussian basis set for carbon consists of uncontracted 8s6p4d orbitals. We used correlation-consistent pseudopotentials with four valence electrons~\cite{ccECP1,ccECP2}. The molecular orbitals are obtained by solving the Kohn–Sham equations on a real-space grid using the local density approximation (LDA)~\cite{lda,Azadi2010}. The AGP ansatz, which is the particle-number-conserving form of the BCS wave function, captures static correlation within a single determinant framework.

Dynamic correlation is incorporated through the Jastrow factor
\begin{equation}
J(\mathbf{R}) = \exp\left( \sum_{i<j} u(\mathbf{r}_i, \mathbf{r}_j) \right),
\end{equation}
where $u(\mathbf{r}_i, \mathbf{r}_j)$ is a two-body correlation function depending on the inter-electron distance $r = |\mathbf{r}_i - \mathbf{r}_j|$, taken as
\begin{equation}
u(r) = \frac{r}{2(1 + \eta r)},
\end{equation}
with $\eta$ a variational parameter~\cite{Fahy90}. The Jastrow orbitals are expanded in an uncontracted Gaussian basis of 4s3p. All variational parameters are optimized by energy minimization using the linear method~\cite{Umrigar2007} and stochastic reconfiguration (SR)~\cite{Sorella98}.

Diffusion Monte Carlo (DMC) calculations are performed with a time step of 0.01 a.u., using 1920 walkers and the locality approximation for the pseudopotential~\cite{Casula2006}. The JAGP wave function is constructed from the JSD ans\"{a}tz with maximal overlap, ensuring continuity of the nodal structure. During wave function optimization, convergence is monitored by maximizing the signal-to-noise ratio while minimizing the energy. All QMC simulations were performed using TurboRVB code \cite{TurboRVB}.

\begin{figure*}
    \centering
    \includegraphics[width=1.0\linewidth]{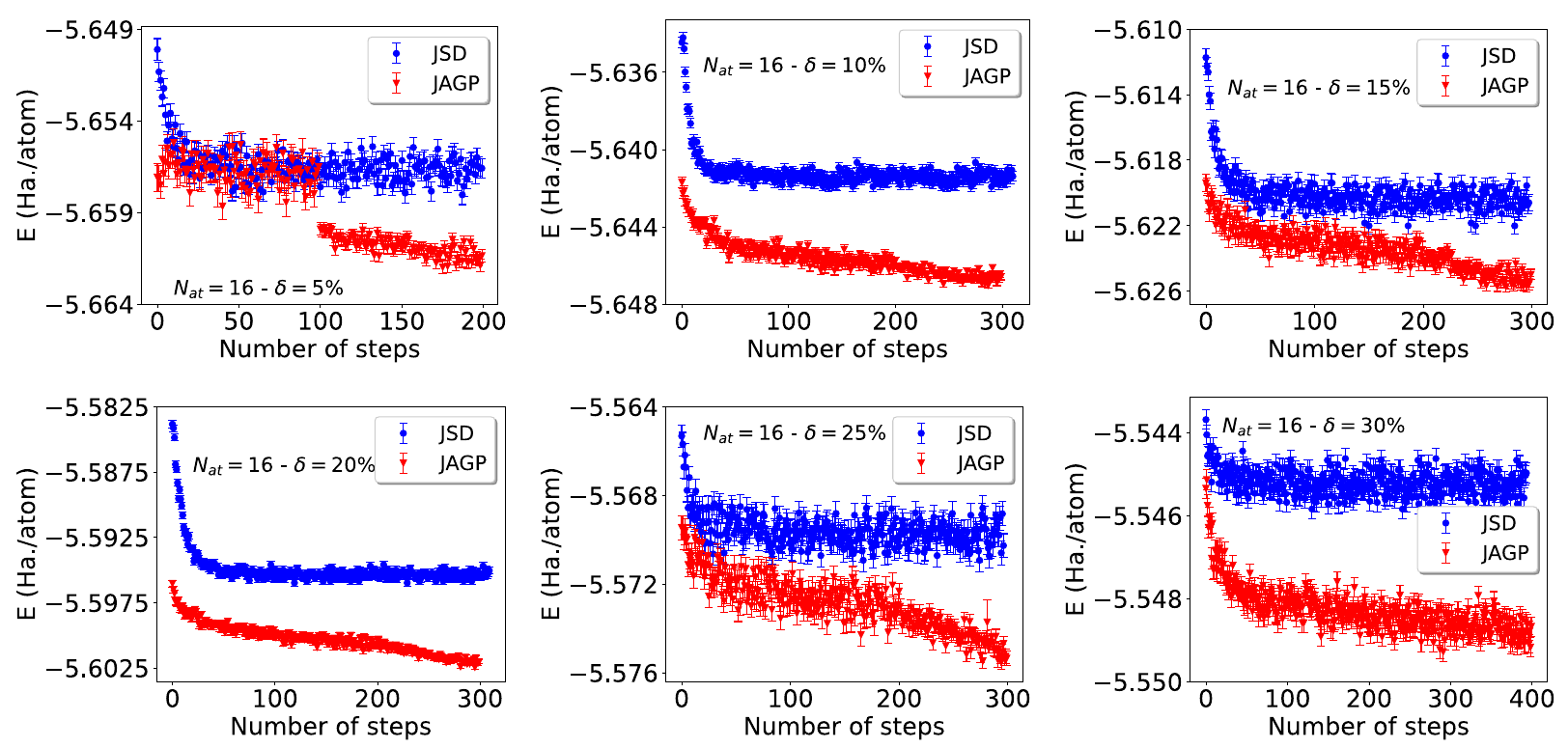}
    \caption{VMC energy as a function of the number of optimization steps computed using JSD and JAGP wave functions for system size N=16.}
    \label{fig:WFOPTN16}
\end{figure*}
\begin{figure*}
    \centering
    \includegraphics[width=1.0\linewidth]{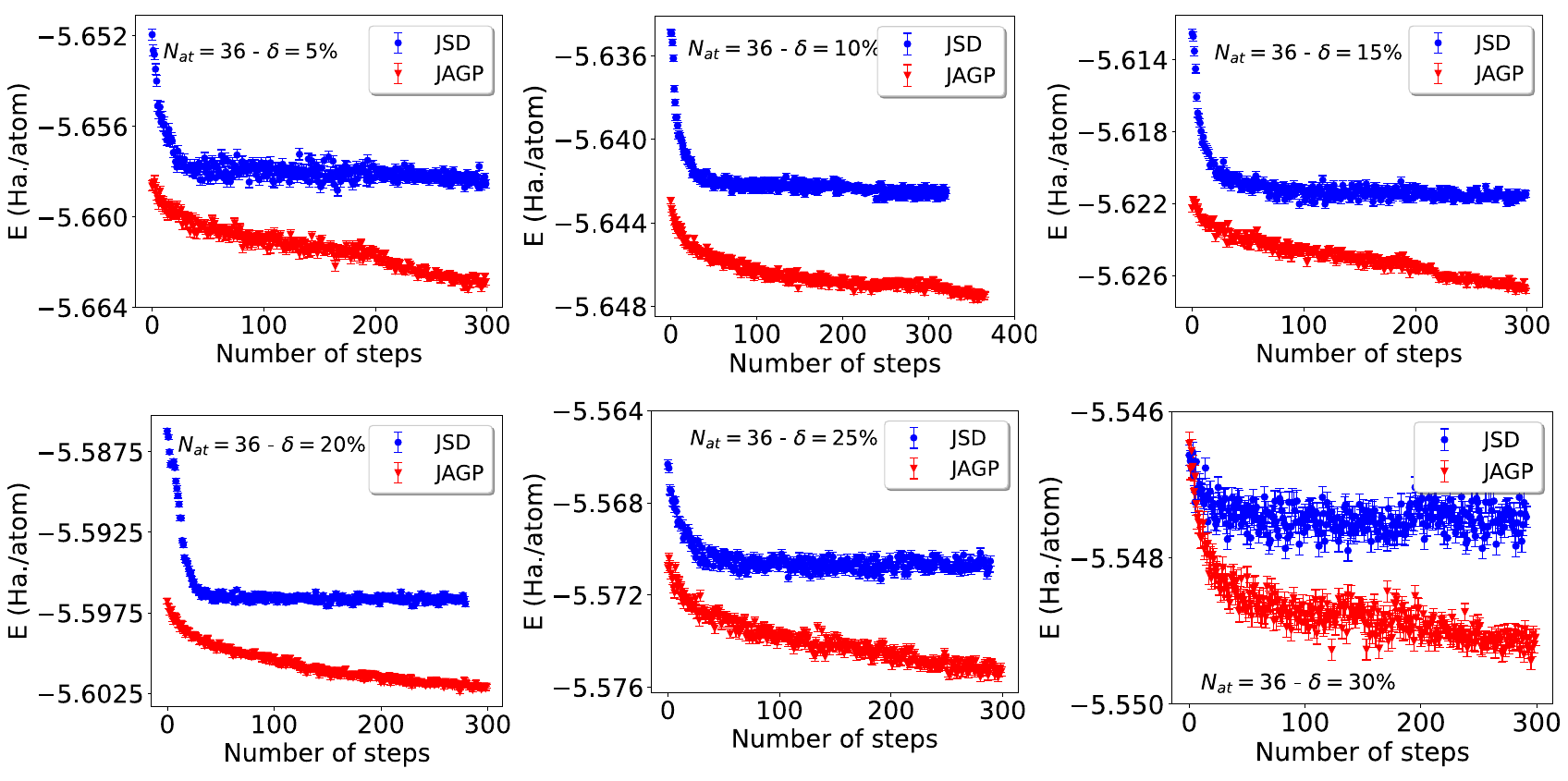}
    \caption{VMC energy as a function of the number of optimization steps computed using JSD and JAGP wave functions for system size N=36.}
    \label{fig:WFOPTN16}
\end{figure*}
\begin{figure*}
    \centering
    \includegraphics[width=1.0\linewidth]{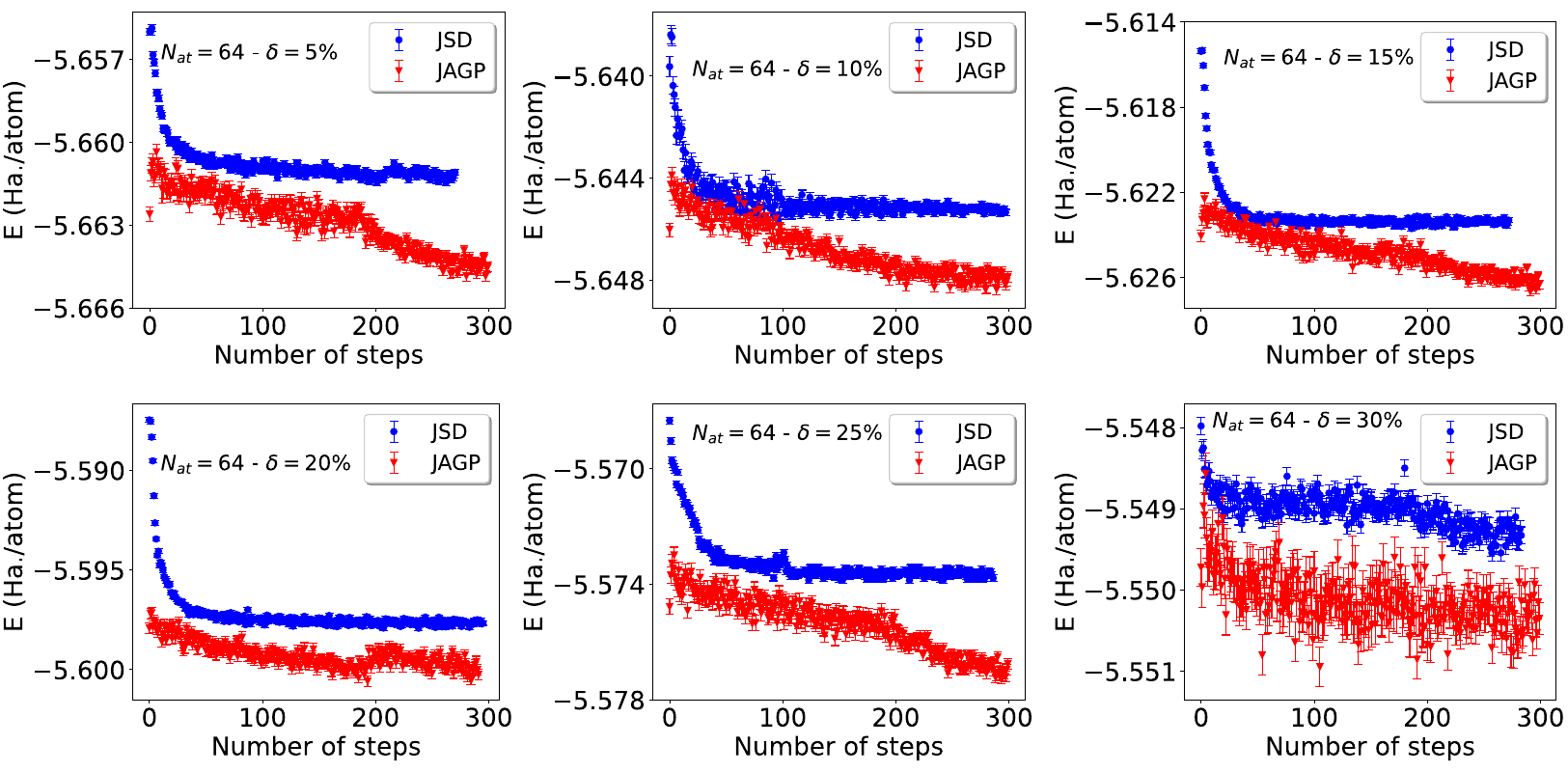}
    \caption{VMC energy as a function of the number of optimization steps computed using JSD and JAGP wave functions for system size N=64.}
    \label{fig:WFOPTN16}
\end{figure*}

\begin{table}[!htb]
    \centering
    \begin{tabular}{|c|cccc|}
    \hline\hline
    $\delta$ & VMC-JSD  & DMC-JSD  & VMC-JAGP & DMC-JAGP  \\
    \hline
    $0\%$  & -90.6322(3) & -90.858(1) & -90.6744(2) & -90.871(1)\\
    $5\%$  & -90.5081(2) & -90.765(1) & -90.5843(2) & -90.792(1)\\
    $10\%$ & -90.2628(2) & -90.534(1) & -90.3475(2) & -90.568(1)\\ 
    $15\%$ & -89.9267(2) & -90.217(1) & -90.0205(2) & -90.256(1)\\ 
    $20\%$ & -89.5269(2) & -89.835(1) & -89.6236(2) & -89.881(1)\\
    $25\%$ & -89.1168(2) & -89.430(1) & -89.2008(2) & -89.461(2)\\
    $30\%$ & -88.7249(6) & -89.029(2) & -88.7793(3) & -89.052(2)\\
    \hline\hline
    \end{tabular}
    \caption{VMC and DMC total energies in Ha obtained using JSD and JAGP WFs using system size with N=16 atoms. The first column lists values of stretch parameter $\delta$. }
    \label{tab:TotE}
\end{table}
\begin{table}[!htb]
    \centering
    \begin{tabular}{|c|cccc|}
    \hline\hline
    $\delta$ & VMC-JSD  & DMC-JSD  & VMC-JAGP & DMC-JAGP  \\
    \hline
    $0\%$  & -204.0647(2) & -204.553(2) & -204.1504(3) & -204.529(2)\\
    $5\%$  & -203.7027(3) & -204.307(2) & -203.8662(3) & -204.358(5)\\
    $10\%$ & -203.1341(3) & -203.751(5) & -203.3099(2) & -203.811(5)\\ 
    $15\%$ & -202.3774(3) & -203.037(3) & -202.5618(2) & -203.106(4)\\ 
    $20\%$ & -201.4784(3) & -202.171(5) & -201.6748(2) & -202.241(5)\\
    $25\%$ & -200.5394(3) & -201.263(4) & -200.6868(3) & -201.325(6)\\
    $30\%$ & -199.7123(2) & -200.268(4) & -199.7707(3) & -200.308(6)\\
    \hline\hline
    \end{tabular}
    \caption{VMC and DMC total energies in Ha obtained using JSD and JAGP WFs using system size with N=36 atoms. The first column lists values of stretch parameter $\delta$. }
    \label{tab:TotE}
\end{table}
\begin{table}[!htb]
    \centering
    \begin{tabular}{|c|cccc|}
    \hline\hline
    $\delta$ & VMC-JSD  & DMC-JSD  & VMC-JAGP & DMC-JAGP  \\
    \hline
    $0\%$  & -363.042(1)  & -363.91(1) & -363.1958(8) & -364.07(1)\\
    $5\%$  & -362.3180(5) & -363.386(8)& -362.5483(7) & -363.62(2)\\
    $10\%$ & -361.2931(5) & -362.39(1) & -361.5508(5) & -362.69(2) \\ 
    $15\%$ & -359.8991(3) & -361.044(8)& -360.1926(4) & -361.36(3)\\ 
    $20\%$ & -358.2506(4) & -359.494(7)& -358.5089(5) & -359.75(3)\\
    $25\%$ & -356.7133(3) & -357.758(6)& -356.9419(5) & -357.93(3) \\
    $30\%$ & -355.1601(3) & -356.326(5)& -355.2442(4) & -356.47(3) \\
    \hline\hline
    \end{tabular}
    \caption{VMC and DMC total energies in Ha obtained using JSD and JAGP WFs using system size with N=64 atoms. The first column lists values of stretch parameter $\delta$. }
    \label{tab:TotE}
\end{table}



\end{document}